# Interfacial thermal conductance between TiO$_2$ nanoparticle and water: A molecular dynamics study

Mahdi Roodbari[1], Mohsen Abbasi[2], Saeed Arabha[1], Ayla Gharedaghi[1], Ali Rajabpour[1,3]*

[1]*Advanced Simulation and Computing Laboratory, Imam Khomeini International University, Qazvin, Iran*

[2]*Departmant of Mechanical Engineering, Tarbiat Modares University, Tehran, Iran*

[3]*School of Nano Science, Institute for Research in Fundamental Sciences (IPM), Tehran, Iran*

**Abstract**

The interfacial thermal conductance (Kapitza conductance) between a TiO$_2$ nanoparticle and water is investigated using transient non-equilibrium molecular dynamics. It is found that Kapitza conductance of TiO$_2$ nanoparticles is one order of magnitude greater than other conventional nanoparticles such as gold, silver, silicon, platinum and also carbon nanotubes and graphene flakes. This difference can be explained by comparing the contribution of electrostatic interactions between the partially charged titanium and oxygen atoms and water atoms to the van der Waals interactions, which increases the cooling time by about 10 times. The effects of diameter and temperature of nanoparticle, surface wettability on the interfacial thermal conductance are also investigated. The results showed that by increasing the diameter of the nanoparticle from 4 to 9 nm, Kapitza conductance decreased slightly. Also, increasing the temperature of the heated nanoparticle from 400 K to 600 K led to thermal conductance enhancement. It has been found that increasing the coupling strength of Lennard-Jones (LJ) potential from 0.5 to 4 caused the increment of the Kapitza conductance about 20%. It is also shown that a continuum model which its input is provided by molecular dynamics can be a suitable approximation to describe the thermal relaxation of a nanoparticle in a liquid medium.

* Corresponding Author: Rajabpour@eng.ikiu.ac.ir

# 1. Introduction

Nanofluid, a homogeneous mixture of nanoparticles and base fluid, has interesting thermal properties [1]. Nowadays, nanofluids possess many applications in industries, such as solar thermal systems, cooling of electronic devices, and pool and flow boiling [2,3,4,5,6]. Thus, it is important to investigate the heat transfer in the proximity of nanoparticles that are surrounded by water. The thermal properties of nanofluids can be affected by heat transfer at the interface of the nanoparticle as a solid and water as a fluid. When a nanofluid is exposed to a thermal gradient, a temperature jump at the interface of the nanoparticle and water can be observed. This temperature jump plays an important role in the thermal resistance between water and nanoparticle. For the first time, in 1941, Kapitza [7] studied heat transfer between a metal and liquid helium, and the concept of interfacial thermal resistance (ITR) or Kapitza resistance was introduced. Furthermore, Keblinski et al. [8] proposed a nanolayer at the interface of solid-liquid for analyzing heat transfer in nanofluids. In recent years, many researchers have studied the ITR and nanolayer to measure their influences on thermophysical properties of nanofluids [9,10].

Both experimental and simulation methods are employed at the atomic scale to investigate ITR between nanoparticles and fluids [11,12]. Because experimental research consumes time and cost, Molecular Dynamics (MD) simulations provide the prediction of properties of materials effectively. MD has been extensively utilized to determine the characteristics of nanofluids, such as thermal conductance and viscosity [13,14]. The interfacial thermal conductance can be calculated by four methods via: transient non-equilibrium molecular dynamics (TNEMD) methods with and without considering temperature gradient in nanoparticle, respectively, steady-state non-equilibrium method (SNEMD) and equilibrium (EMD) method [10]. Several theoretical studies have been conducted to investigate heat transfer at the interfaces of different particles using these methods. Thermal conductance between water and surfaces of gold, silver, silicon, and copper in nanochannels was studied employing SNEMD simulation [15]. It was observed that the interaction strength between solid and fluid at the metallic interfaces influences the water molecules' behavior, and as a result, ITR is affected. The heat conduction in nanochannels was studied using the EMD method [16]. It was showed that when the heat transferred at the interface of liquid-solid, a temperature jump occurred at the interface, known as Kapitza resistance. The effect of wetting of surface on the strength of solid-liquid interactions was studied [17], and the results showed that exponential and power-law forms of Kapitza strength functions are related to non-wetting and wetting liquids, respectively. Also, the effects of surface wettability and thermal oscillation frequency on Kapitza length were examined [18,19]. It was found that by increasing the wettability of the surface and decreasing the thermal oscillation frequency, the Kapitza length reduces.

The mechanism of heat transfer with high heat flux at the interface of gold nanoparticle surrounded by water/octane was investigated using the SNEMD method [20,21]. A nanoparticle and a planar wall were heated and enclosed by water. The results indicated that the phase of water around the heated nanoparticle did not change, whilst water around the heated planar wall converted to a layer of vapor, and this reduced the amount of heat flux. SNEMD and continuum theory were utilized to investigate the effect of Kapitza resistance on the rate of evaporation of water droplet at a heated surface [22,23]. It was showed that thermodynamics conditions can determine the influence of Kapitza resistance on heat flux.

Interfacial thermal conductance between graphene sheets and solid/liquid octadecane ($C_{18}H_{38}$) was studied using NEMD [24]. It was found that the interfacial thermal conductance of solid paraffin-graphene is higher than liquid paraffin-graphene. The dependence of ITR on heat flux direction in hybrid graphene-graphane nanoribbons was investigated [25]. It was demonstrated that Kapitza resistance strongly depends on the imposed direction of heat flux. The Kapitza resistance of CNT and graphene flake was investigated employing the SNEMD method [26]. It was concluded that the interaction strength of carbon/water does not depend on the nanolayer around the nanoparticle. The effect of polymer coating on the ITR of gold-water was studied utilizing EMD simulation [27]. The results showed that coating polymer led to increasing the interfacial thermal conductance. $TiO_2$ nanoparticles and coatings are one of the most widely used materials in the nanotechnology industry. This material is very popular due to its hydrophobic property. Also, due to the partial charges on titanium and oxygen atoms, it has polar interactions with water molecules, which can distinguish the rate of heat transfer from these nanoparticles to water compared to other similar nanoparticles that only interact with van der Waals forces.

In the present study, the interfacial thermal conductance between $TiO_2$ nanoparticle and surrounding water is calculated using TNEMD simulation. The parameters that affect the thermal conductance and thermal relaxation time of the nanoparticle are investigated: diameter of nanoparticle from 4 to 9 nm, nanoparticle temperature range from 400 K to 600 K, surface wettability and Coulombic interatomic potential contribution. Finally, to investigate the effect of water molecules movement on thermal conductance, the finite volume method is used to solve the equations of heat conduction between the surrounded nanoparticle and water and the results are compared with MD simulation results.

## 2. Simulation details

### 2.1. Simulation setup

In the present study, large-scale atomic/molecular massively parallel simulator (LAMMPS [28]) is employed for molecular dynamic simulations. In our model system, a titanium dioxide nanoparticle is immersed in water. The shape of the nanoparticle is spherical with diameters of 4 to 9 nm. Fig. 1 shows the geometry of a titanium dioxide nanoparticle with a diameter of 4 nm which is surrounded by water molecules. The water molecules are distributed randomly using Packmol [29] package.

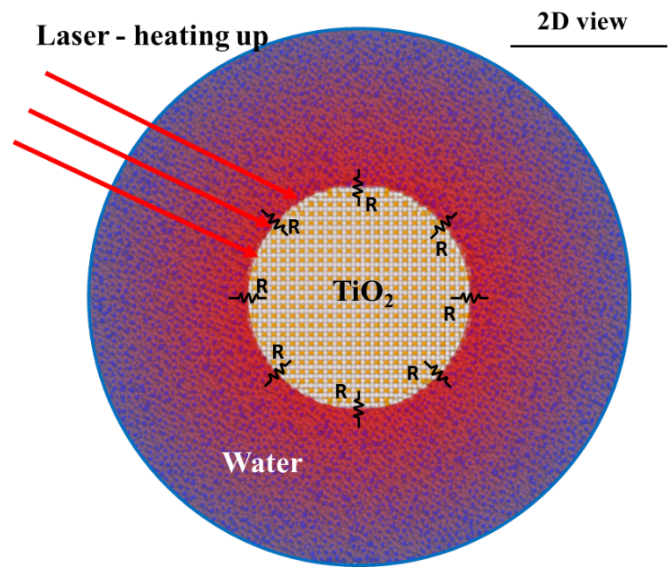

Fig. 1. Molecular dynamics model of $TiO_2$ nanofluid: atomistic structures of $TiO_2$ nanoparticle with a diameter of 4 nm, and surrounding water molecules. R (=1/G) represents the interfacial thermal resistance between the nanoparticle and water.

To describe the intermolecular forces between water particles, the TIP4P/2005 force-field is utilized [30]. In this model, the partial charges of hydrogen and oxygen are 0.5564$e$ and -1.1128$e$, respectively. The angle between the bonds of a water molecule is equal to 104.5°, and the length of the bond of oxygen-hydrogen is 0.96 Å. The mass of the hydrogen atom is assumed to be negligible, therefore, the interaction between hydrogen atoms in a water molecule is neglected. The potential energy, which determines the forces acting on the atoms, includes Lennard-Jones (LJ) potential and long-range Coulombic interactions [31]. The sum of these potentials is used for calculating the interactions between atoms of a water molecule is shown in Equation (1).

$$U_{ij}(r_{ij}) = \sum_i^a \sum_j^b \frac{1}{4\pi\epsilon_0} \frac{q_i q_j}{r_{ij}} + 4\varepsilon_{ij}\left[\left(\frac{\sigma_{ij}}{r_{ij}}\right)^{12} - \left(\frac{\sigma_{ij}}{r_{ij}}\right)^6\right] \qquad (1)$$

where $\epsilon_0$ represents the dielectric constant of vacuum, r$_{ij}$ is the distance between i and j atoms, and $q_i$ and $q_j$ are the electric charge of atoms *i* and *j*, respectively. $\varepsilon_{ij}$ and $\sigma_{ij}$ are energy and distance units of LJ potential.

To simulate the TiO$_2$ nanoparticle, its ionic structure should be addressed. There are three crystalline phases of TiO$_2$ nanoparticle, rutile, anatase, and brookite. The spatial group, crystal dimension, crystal structure, and thermodynamic stability are different in each crystalline phase. The crystalline phase greatly depends on the size of the nanoparticle. Anatase is the dominant crystalline phase in the small size of nanoparticle (less than 14 nm) [32]. Materials Studio software is used for generating the anatase phase of TiO$_2$ nanoparticle, and its parameters are taken from experimental results which are shown in Table 1 [33]. To validate the created atomic structure, the bulk density of TiO$_2$ nanoparticle is calculated at 30°C without the presence of water, that is obtained 3.91 g/cm$^3$ which is in good agreement with experimental result [34].

Table 1. Parameters of the unit cell for simulation of TiO$_2$ nanoparticle in the anatase phase [35]

| Crystal structure | a (Å) | b (Å) | c (Å) |
|---|---|---|---|
| Tetragonal | 3.785 | 3.785 | 9.513 |

The Matsui-Akaogi force-field [42] has been employed for simulating TiO$_2$ nanoparticle [36,37,38,39,40]. The electrostatic force determines the interactions between the atoms of titanium and oxygen in the TiO$_2$ nanoparticle. The partial charges of titanium and oxygen atoms are 2.19 e and 1.098 e, respectively. In order to calculate van der Waals forces between the particles, the potential of Buckingham is utilized [41]. Table 2 represents the coefficients of Buckingham potential of the Matsui-Akaogi force-field (Equation (2)).

$$U(r_{ij}) = A_{ij} exp\left(\frac{-r_{ij}}{\rho_{ij}}\right) - \frac{B_{ij}}{r_{ij}^6} + \frac{q_i q_j}{r_{ij}} \tag{2}$$

Table 2. Coefficients of Matsui-Akaogi force-field [42]

| Type of interaction | A$_{ij}$ (kcal/mol) | B$_{ij}$ (kcal/mol$^{-1}$ Å$^6$) | $\rho_{ij}$ (Å) |
|---|---|---|---|
| Titanium- Titanium | 717650 | 121.1 | 0.155 |
| Titanium-Oxygen | 391050 | 290.42 | 0.195 |
| Oxygen-Oxygen | 271720 | 696.95 | 0.235 |

The LJ and long Coulombic potentials are employed for interactions between water molecules and TiO$_2$ nanoparticle [43]. The parameters of LJ potential for different types of atoms are calculated using a combination formula proposed by Lorentz-Berthelot which are shown in Equations (3) and (4) [44].

$$\sigma_{pf} = \frac{(\sigma_{pp} + \sigma_{ff})}{2} \tag{3}$$

$$\varepsilon_{pf} = \sqrt{\varepsilon_{pp}.\varepsilon_{ff}} \tag{4}$$

where p and f imply non-bonding atoms in the system. Table 3 shows the values of parameters of LJ potential for different atomic interaction.

Table 3. The coefficients of LJ potential in the current study [43,44]

| Interaction | $\sigma(\text{Å})$ | $\varepsilon(Kcal/mol)$ |
|---|---|---|
| H-H | 0 | 0 |
| H-$O_w$ | 0 | 0 |
| $O_w$-$O_w$ | 3.16 | 0.182 |
| Ti-$O_w$ | 2.99 | 0.056 |
| $O_T$-$O_w$ | 3.54 | 0.097 |
| Ti-H | - | 0 |
| $O_T$-H | - | 0 |

$O_T$: O atom of $TiO_2$, $O_W$: O atom of $H_2O$

## 2.2. Simulation processes

The technique of Ewald summation and the method of cut-off are utilized to compute the electrostatic and short-range forces, respectively. The time step in all simulations is considered 1 fs. To remove close contacts and avoiding high collisions of potential energy, minimization has been done. First, the system is equilibrated for 10 ps in NPT ensemble, and temperature and pressure of the system are controlled at 300 K and 1 atm. Langevin thermostat is used to keep the temperatures of nanoparticle and water at 300 K. Then the temperature of nanoparticle increased to 400 K, while the temperature of water is fixed at 300 K. The NVE ensemble is used in the process of thermal relaxation of the nanoparticle, and heat is dissipated from surrounded nanoparticle to the water which its temperature is still controlled at 300 K. The variations of the temperatures of $TiO_2$ nanoparticle with a diameter of 6 nm and water during transient non-equilibrium MD simulation are shown in Fig. 2. As expected, the temperature of the nanoparticle decreases toward the temperature of water.

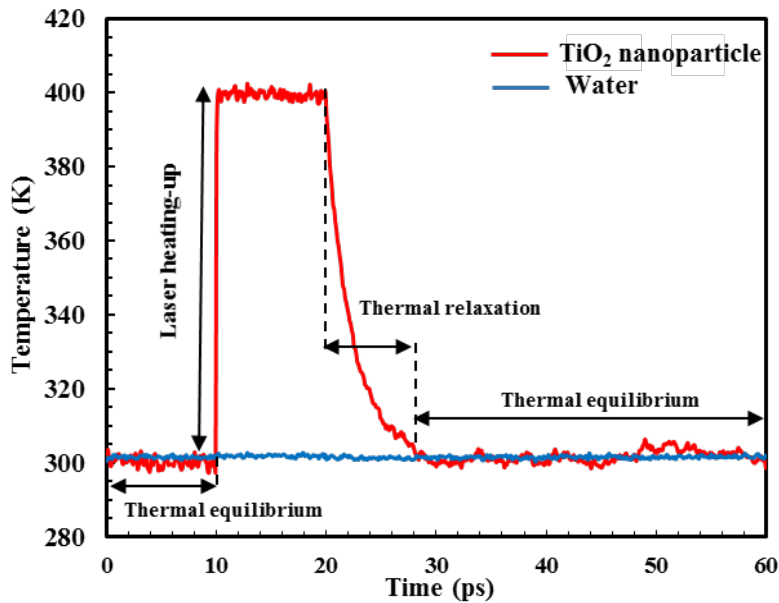

Fig. 2. Temperature variations of TiO$_2$ nanoparticle with a diameter of 6 nm and water during the process of MD simulation. The temperature of water is fixed at 300 K using Langevin thermostat.

In order to assure that the Langevin thermostat keeps the temperature of water molecules in the cooling time of nanoparticle at 300 K, the temperatures of water at three shells around the nanoparticle were investigated. Three spherical shells are concentric with the thickness of t=1 nm adjacent to the nanoparticle. Fig. 3 represents the temperatures of water molecules of these shells. It can be observed from Fig. 3 that the water temperature fluctuated around 300 K in the cooling time of the nanoparticle, which indicates the thermostat worked properly.

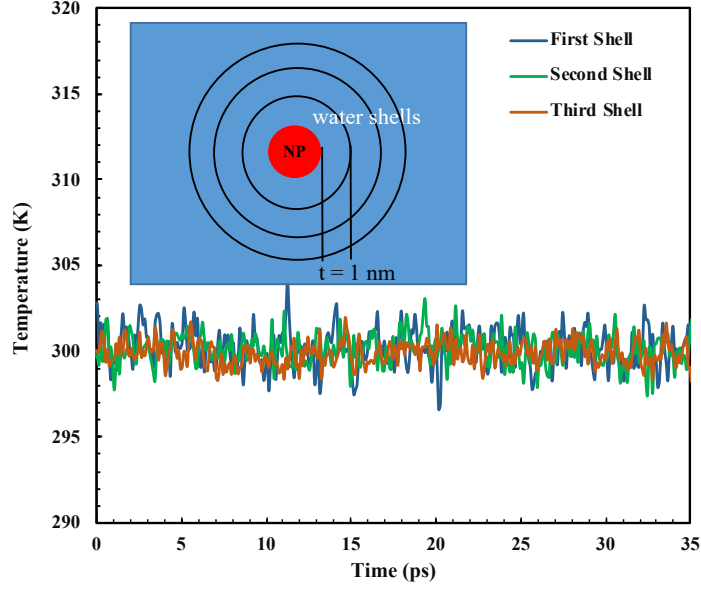

Fig. 3. Temperature of the different water spherical shells with thickness of t = 1 nm around the TiO$_2$ nanoparticle during the thermal relaxation of the nanoparticle. The water temperature of these shells fluctuates around 300 K because of the applied thermostat.

## 3. Interfacial thermal conductance

When the heat flows through an interface of two bodies, a temperature jump at the interface occurs known as Kapitza resistance. This happens for two reasons: 1) two surfaces do not contact completely 2) the vibrational characteristics of two materials differ [45]. Interfacial thermal resistance (Kapitza resistance) can be calculated using Fourier's law [46]:

$$Q = \frac{\Delta T}{R} \tag{5}$$

where Q is the heat flux (W/m$^2$) across the interface of two bodies, $\Delta T$ is temperature jump (K), and $R$ is Kapitza thermal resistance (m$^2$ K/W). The interfacial thermal conductance (G) is the inverse of Kapitza thermal resistance as follows:

$$G = \frac{1}{R} \tag{6}$$

### 3.1. Kapitza resistance calculation method

The difference between temperatures of TiO$_2$ nanoparticle and water causes energy exchanges in the system. During this process, the temperature and total energy of the surrounded nanoparticle are calculated and recorded. Kapitza resistance can be calculated using the following Equation (7) which is derived from the total energy conservation for the system:

$$\frac{\partial E_t}{\partial t} = -GA(T_{np} - T_W) = -\frac{A(T_{np} - T_W)}{R} \qquad (7)$$

where $\frac{\partial E_t}{\partial t}$ is internal energy variation of the nanoparticle with respect to the time and $GA(T_{np} - T_W)$ is the heat dissipated into the water. $A$ is the surface area of heat transfer which is the surface area of the nanoparticle, and $T_{np}$, $T_W$ are the temperatures of nanoparticle and water, respectively. The measured temperatures of nanoparticle and water are uniform. Equation (7) should be integrated, and $R$ is assumed to be a constant value during the cooling, as follows:

$$E_t = \frac{-A}{R} \int_0^t (T_{np} - T_W) dt + E_0 \qquad (8)$$

where $E_0$ [J] is the initial total energy of TiO$_2$ nanoparticle just before starting the thermal relaxation of the nanoparticle. By knowing the temperature difference between nanoparticle and water in each time step and using the trapezoidal rule for numerical integration, the value of $\int_0^t (T_{np} - T_W) dt$ can be computed. Fig. 4 shows the change of $E_t$ with respect to $\int_0^t \Delta T dt$. A linear curve is fitted and its slope is calculated. Therefore, Kapitza resistance between TiO$_2$ nanoparticle and water can be calculated.

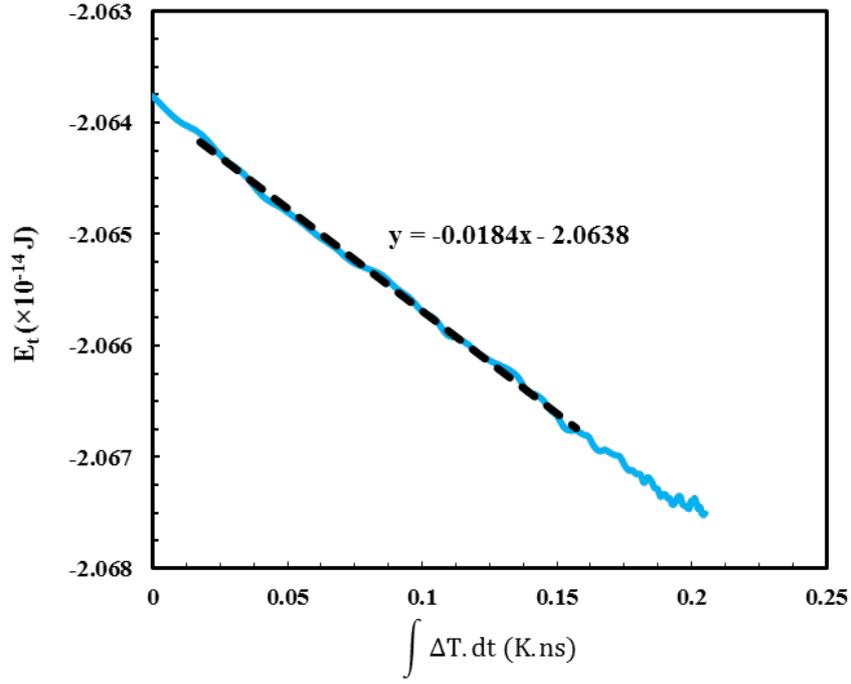

Fig. 4. Variation of $E_t$ versus $\int_0^t \Delta T \, dt$ for TiO$_2$ nanoparticle with a diameter of 6 nm

The constant value assumption for R is confirmed due to the linear association of $E_t$ and $\int_0^t \Delta T \, dt$. The fluctuation at the end of curve is not considered, due to negligible amount of temperature difference between water and nanoparticle. Also, obtained results at initial time steps are not considered, because at these time steps, kinetic energy converts to potential energy, and acquired data is not correct sufficiently, therefore, they are neglected in the curve fitting [47].

## 4. Results and discussion

In this section, the influences of the diameter, wettability, and temperature of nanoparticle on thermal conductance are examined. Also, the effect of Coulombic interatomic potential on the cooling process of nanoparticle is investigated.

### 4.1. Effect of nanoparticle diameter

The decay of the temperature of the nanoparticle with different diameters in the relaxation period is shown in Fig. 5. It can be seen that the TiO$_2$ nanoparticle with a diameter of 9 nm needs more time than the diameter of 4 nm to reach the equilibrium temperature. This is attributed to the initial energy of the nanoparticle, which is higher for bigger nanoparticles, and as a result, the thermal relaxation time increases. The higher the diameter of the nanoparticle, the slower the temperature decay.

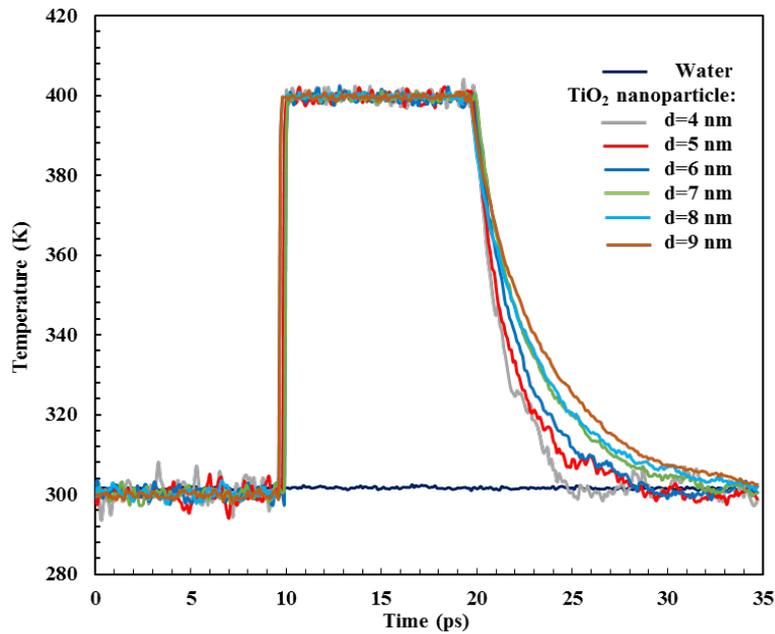

Fig. 5. Temperature variations of $TiO_2$ nanoparticle with diameters of 4-9 nm during the process of MD simulation

Fig. 6 shows the value of interfacial thermal conductance (G) for $TiO_2$ nanoparticle with various diameters. As it can be observed, the thermal conductance decreases by increasing the diameter of nanoparticle from 4 nm to 9 nm. Interfacial thermal conductance is determined by nanolayer, which is a thin layer of liquid formed at the liquid/solid interface, and is a thermal bridge between nanoparticle and water. According to [40,48], the density of nanolayer increases by increasing the diameter of the nanoparticle. If the diameter of the nanoparticle goes to infinity, the density of nanolayer reaches to its maximum amount but not the infinity. The concentrated density of nanolayer acts as a barrier between the nanoparticles and water, reducing interfacial thermal conductance. As it can be seen from Fig. 6, the slope of downward trend of reduction of interfacial thermal conductance is decreasing. Therefore, if the diameter of the nanoparticle goes to infinity, interfacial thermal conductance has the minimum amount. The values of the interfacial thermal conductance are the average of the 5 simulation results, so, the error bar indicates the minimum and maximum values of the results.

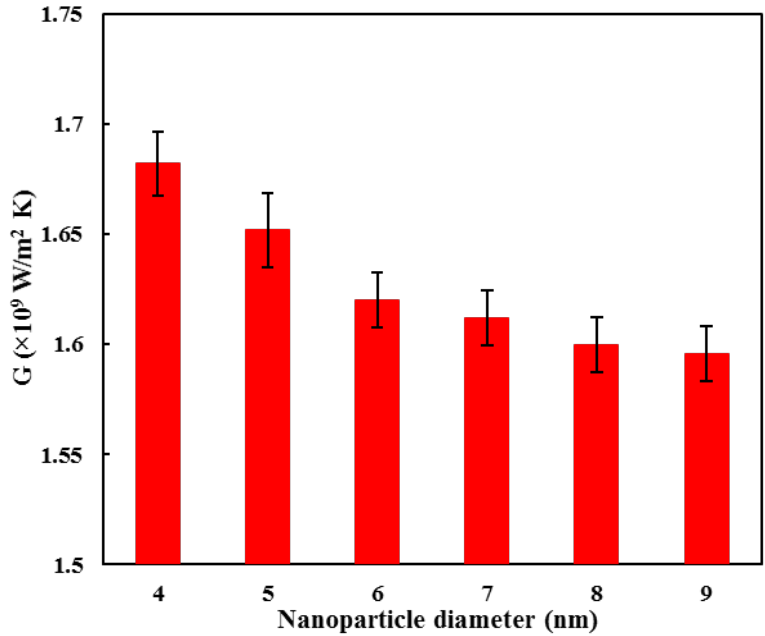

Fig. 6. Thermal conductance between nanoparticle and water with respect to the diameter of $TiO_2$ nanoparticle

*4.2. Effect of nanoparticle temperature*

Fig. 7 illustrates the effect of the temperature of heated nanoparticle on the thermal relaxation time. Using a Langevin thermostat, the temperature of $TiO_2$ nanoparticle with a diameter of 6 nm increased to 400-550 K. It can be observed that time of reaching the equilibrium temperature during thermal relaxation time is shorter for nanoparticle with less temperature. The heated nanoparticle with temperature of 400 K reached the temperature of water in 27 ps, while it was 32 ps for the nanoparticle with temperature of 550 K. Therefore, the thermal relaxation time depends on the temperature of heated nanoparticle. The effect of increasing the temperature of nanoparticle on thermal conductance is discussed in the next section.

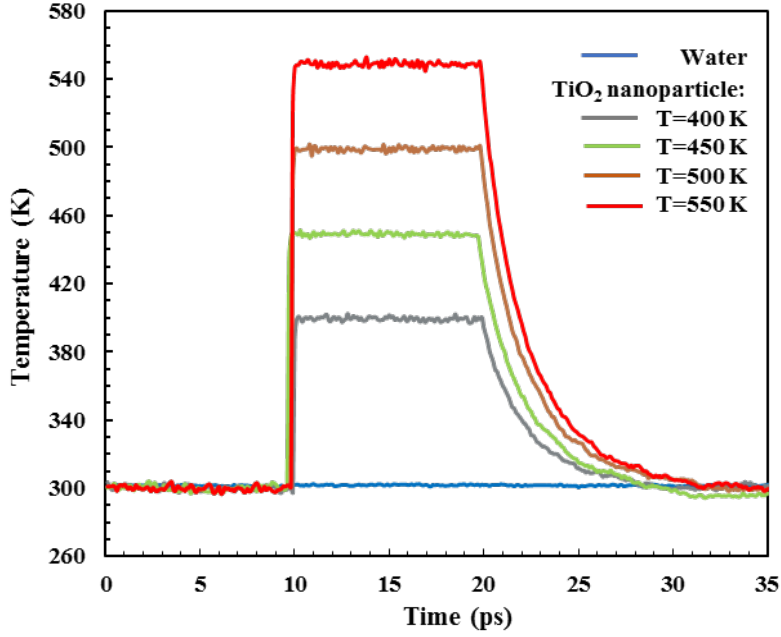

Fig. 7. The effect of variation of TiO$_2$ nanoparticle temperature (in heating-up stage) on the thermal relaxation process.

*4.3. Effect of surface wettability*

As discussed earlier, the Lennard-Jones potential plays an important role in the interaction between water molecules and TiO$_2$ nanoparticle. The parameter $\alpha$ is defined as the strength of the particles' interaction, which is used to the investigation of surface wettability, as follows:

$$\alpha = \frac{\varepsilon_{pf}}{\varepsilon_{pf_0}} \tag{9}$$

where $\varepsilon_{pf}$ and $\varepsilon_{pf_0}$ are the energy parameters of Lennard-Jones interatomic potential, which are related to the interactions between atoms of nanoparticle and base fluid. The former parameter indicates the initial conditions of the interaction between particles of TiO$_2$ nanoparticle and water, and the latter one represents changes in the coefficients of interaction. The different values of $\alpha$ (0.5, 1, 2, 3, and 4) and various nanoparticle temperatures are employed to investigate the effect of surface wettability of the nanoparticle on thermal conductance. Fig. 8 shows the effect of surface wettability and temperature of TiO$_2$ nanoparticle with a diameter of 5 nm on thermal conductance. It can be seen that at the same temperature of the nanoparticle, by increasing the value of $\alpha$, thermal conductance rises. This is related to the better vibrational correlation between surface of nanoparticle and water. When the strength of interaction between solid and liquid increases the thermal conductance rises [49].

The temperature of heated TiO$_2$ nanoparticle is studied at 400-600 K to investigate the influence of heated nanoparticle temperature on thermal conductance using MD simulations. The variations of thermal conductance of TiO$_2$ nanoparticle and water in different nanoparticle temperatures in heating-up stage are represented in Fig. 8. As it can be observed, by increasing the temperature of nanoparticle from 400 to 600, thermal conductance between water and nanoparticle increases.

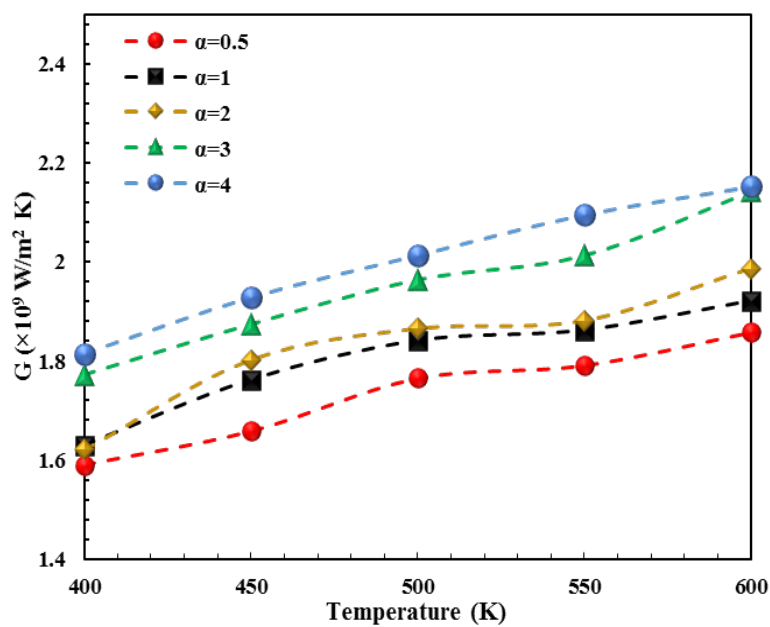

Fig. 8. Variation of thermal conductance (G) between TiO$_2$ nanoparticles with a diameter of 5 nm and water due to temperature and surface wettability of nanoparticle

By using Fourier's transform of the velocity autocorrelation function (VACF) of hydrogen and oxygen atoms, the vibrational spectra (DOS) of water molecules in the presence of TiO$_2$ nanoparticle with a diameter of 6 nm is shown in Fig. 9. The VACF can be calculated using Eq. 10.

$$VACF(t) = <V(0).V(t)> \quad (10)$$

where V(0) and V(t) are the velocities of atoms of hydrogen, oxygen, and titanium at t = 0 and t = t and < > denotes the ensemble averaging. The DOS overlap between the sides of the interface of water molecules and $TiO_2$ nanoparticle at various temperature is calculated as follows:

$$S = \frac{\int DOS_1(\omega)DOS_2(\omega)d\omega}{(\sqrt{DOS_1^2(\omega)d\omega})(\sqrt{DOS_2^2(\omega)d\omega})} \quad (11)$$

where 1 and 2 signify the sides of water and nanoparticle, respectively, and $\omega$ is the angular frequency. The value of $S$ is variable between 0 and 1, and maximum overlap (S=1) occurs when the both sides of the structure have the same vibrational distribution. In the structure of $TiO_2$ nanoparticle surrounded with water molecules at the temperatures of 300 and 400, the $S$ is calculated as 0.76 and 0.79, respectively. The increasing of the overlap between two spectra by temperature rising could justify the interfacial thermal conductance behavior with respect to the temperature variations.

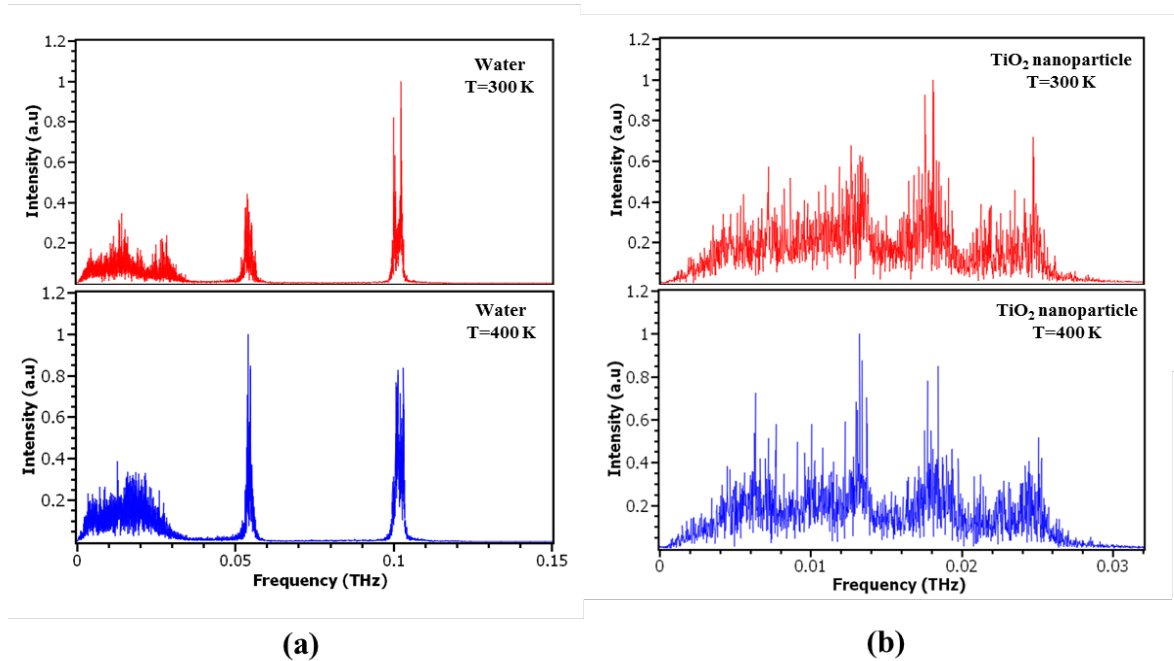

(a)      (b)

Fig. 9. Vibrational spectra of (a) water and (b) $TiO_2$ nanoparticle using Fourier's transform of the velocity autocorrelation function of oxygen, hydrogen, and titanium atoms at temperatures of 300 K and 400 K

*4.4. Effect of Coulombic potential contribution*

As mentioned earlier, the interactions between water atoms and $TiO_2$ nanoparticle follows the potential functions of Lennard-Jones and Coulomb. To investigate the contribution of the Coulombic interatomic potential on the thermal conductance, it is assumed that the Lennard-Jones is the only potential in the interaction between water and $TiO_2$ nanoparticle, and the Coulombic interatomic potential is not considered. Fig. 10 represents the thermal relaxation time of a $TiO_2$ nanoparticle with a diameter of 6 nm with and without considering the Coulombic interatomic potential. It can be observed that when both LJ and Coulombic interatomic potentials are considered for interaction between particles, the time of reaching equilibrium temperature is by far less than when only LJ potential is used. This indicates the large impact of Coulombic interatomic potential on the thermal relaxation time and thermal conductance. The values of thermal conductance between water and $TiO_2$ nanoparticle with diameter of 6 nm with and without considering Coulomb energy potentials are 1620 MW/m$^2$K and 17 MW/m$^2$K, respectively. The reason for this phenomenon is that the electrostatic interactions are stronger than van der Waals interactions, therefore the contribution of transmitted energy through Coulomb potential is greater than that of LJ potential. We are aware that the Coulombic interatomic potential affects the structure of water at the interface such as adsorption density of water molecules at the solid surface and orientation of water molecules [50], but the only way to investigate the effects of LJ and Coulombic interatomic potentials on interfacial thermal conductance is to consider each potential separately. Also, Fig. 10 shows the relaxation time of silver (Ag) nanoparticle with diameter of 1.7 nm surrounded in water [10] compared with relaxation time of $TiO_2$ nanoparticle. It can be found that the relaxation time of silver nanoparticle is close to that of $TiO_2$ nanoparticle when both LJ and Coulombic interaction potentials are considered for interaction between particles of $TiO_2$ nanoparticle and water, while there is a big difference between the relaxation time of these nanoparticles when only LJ interatomic potential is utilized for interaction between $TiO_2$ nanoparticle and water. The noteworthy point here is that due to the high LJ interaction strength for silver/water ($\varepsilon =$ 1.21 $\frac{kJ}{mol}$) compared to the LJ interaction of TiO$_2$/water (Table 3), the interfacial thermal conductance between silver and water is also high.

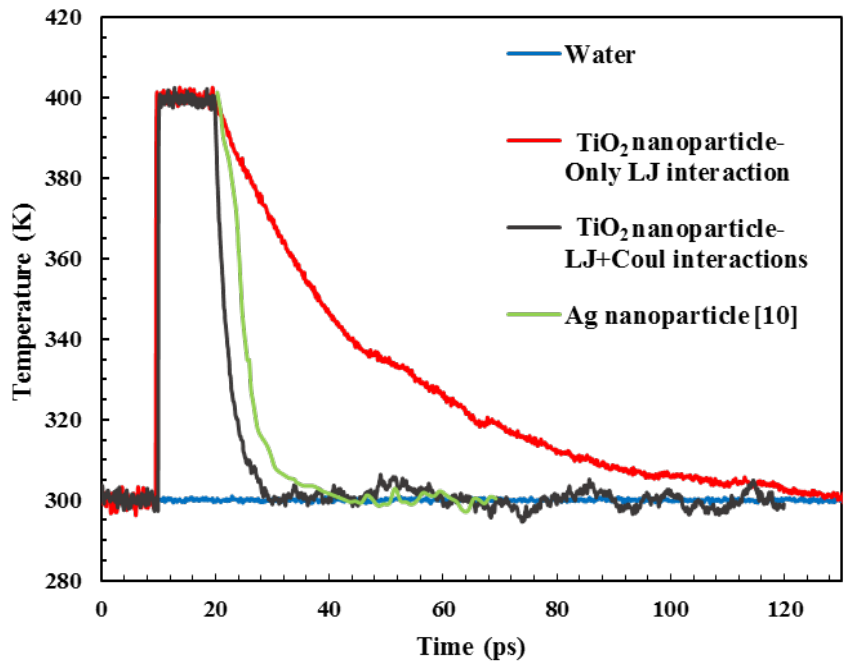

Fig. 10. Thermal relaxation of TiO$_2$ nanoparticle with a diameter of 6 nm when only LJ potential is considered compared to when both LJ and Coulombic potentials are utilized. Also, thermal relaxation time of Ag nanoparticle with diameter of 1.7 nm in water [10] is shown.

## 5. Comparison between the interfacial thermal conductance between different materials

Table 4 shows a comparison between the thermal conductance of TiO$_2$ nanoparticle and nanoparticles of gold, silver, α-Al$_2$O$_3$, CNT, and flat surface of copper, silicon, and platinum at the interface of different liquid types. It can be seen that interfacial thermal conductance of TiO$_2$ nanoparticle is by far more than the others. This is due to the impact of strong electrostatic interactions between the TiO$_2$ nanoparticle and water which is described by Coulombic interatomic potential.

Table 4. Interfacial thermal conductance between different materials

| Solid Type | Liquid type | Interface type | G (MW/m² K) | Ref. |
| --- | --- | --- | --- | --- |
| $TiO_2$ | water | nanoparticle (d=4 nm) | 1682 | Present study |
| Silver (Ag) | water | nanoparticle (d=1.7 nm) | 460 | [10] |
| Gold (Au) | water | nanoparticle (d=3 nm) | 215 | [51] |
| Platinum (Pt) | water | Flat surface | 150 | [15] |
| Silicon (SI) | water | Flat surface | 70.5 | [15] |
| Copper (Cu) | water | Flat surface | 41.4 | [15] |
| CNT | water | nanotube (d=1.1 nm) | 26.4 | [26] |
| α-$Al_2O_3$ | supercritical $CO_2$ | nanoparticle (d=4 nm) | 16.67 | [52] |
| α-$Al_2O_3$ | gaseous $CO_2$ | nanoparticle (d=4 nm) | 6.25 | [52] |

## 6. Comparison between MD and continuum conduction model

In Section 2.2, the nanoparticle cooling process is investigated using the TNEMD method. In this section, we estimate the thermal relaxation process using the continuum conduction model whereas the continuum model inputs are fed from MD outputs. Thus, the effect of motion of water molecules (advection) on the heat transfer mechanism between nanoparticles and water, which has received less attention, is investigated. In MD simulations, water molecules have movements, while in the continuum conduction model, water is considered as a static material [53]. The equations of transient heat conduction model for surrounded $TiO_2$ nanoparticle and water are as follows:

Nanoparticle:

$$T_{rr}^{np} + \frac{2}{r} T_r^{np} = \frac{1}{\alpha^{np}} T_t^{np}$$

$$T_r^{np}(r,t)|_{r=0} = 0$$

$$k_N T_r^{np}(r,t)|_{r=R} = G(T^W(R,t) - T^{np}) \quad (11)$$

$$T^{np}(r,0) = T_i$$

Surrounding water:

$$T_{rr}^W + \frac{2}{r}T_r^W = \frac{\rho^W c_p^W}{k^W}T_t^W$$

$$k^W T_r^W(r,t)|_{r=R} = G(T^W(R,t) - T^{np})$$

$$T^W(r \to \infty, t) = T_0^W \quad (12)$$

$$T^W(R,0) = T_0^W$$

where $T^{np}$ and $T^W$ are the temperatures of nanoparticle and water, respectively, R is radius of nanoparticle, $\alpha^{np}$ is thermal diffusivity of nanoparticle, $k^W$ is thermal conductivity of water, $\rho^W$ is density of water, and $c_p^W$ is the specific heat capacity of water. $T_i$ is equal to 400 K, which is initial temperature of TiO$_2$ nanoparticle, r is the distance from the center of nanoparticle, and $T_0^W$ is the temperature of water at a far distance of r=10 nm, and it is assumed to be 300 K.

The thermo-physical properties of the surrounded TiO$_2$ nanoparticle and water are needed to solve the coupled equations of (11) and (12). For this purpose, internal resistance in nanoparticle is neglected according to lumped model [10]. Based on TIP4P/2005 of MD simulation, the specific heat capacity and thermal conductivity of water are obtained 4.2 kJ/kg

K and 0.8 W/m K, respectively [54]. Using TNEMD calculations, the interfacial thermal conductance between water and $TiO_2$ nanoparticle with a diameter of 9 nm is acquired G = 1596 MW/m$^2$ K. The density of bulk water is assumed to be $\rho^W$=997 kg/m$^3$. However, the density of water in the vicinity of the nanoparticle is different from other parts [**Error! Bookmark not defined.**]. Here, for the simplicity, the density of water is considered to be the same throughout the bulk.

Fig. 11 shows the temperature of model resulted from the numerical solution of Equations of (11), (12) using finite volume method (FVM) and that is directly computed from MD simulation with the same geometry and boundary conditions. It is evident that thermal relaxation time of nanoparticle using MD simulations is close to continuum conduction model. The slight difference between two curves is due to the advection of water molecules which influences the cooling process of the nanoparticle in MD simulation slightly faster than that of continuum model. Thus, the thermal relaxation of a nanoparticle in a liquid could be approximated by a continuum conduction model that its inputs (interfacial thermal conductance between the nanoparticle and liquid, density and thermal conductivity of the liquid) come from a MD simulation.

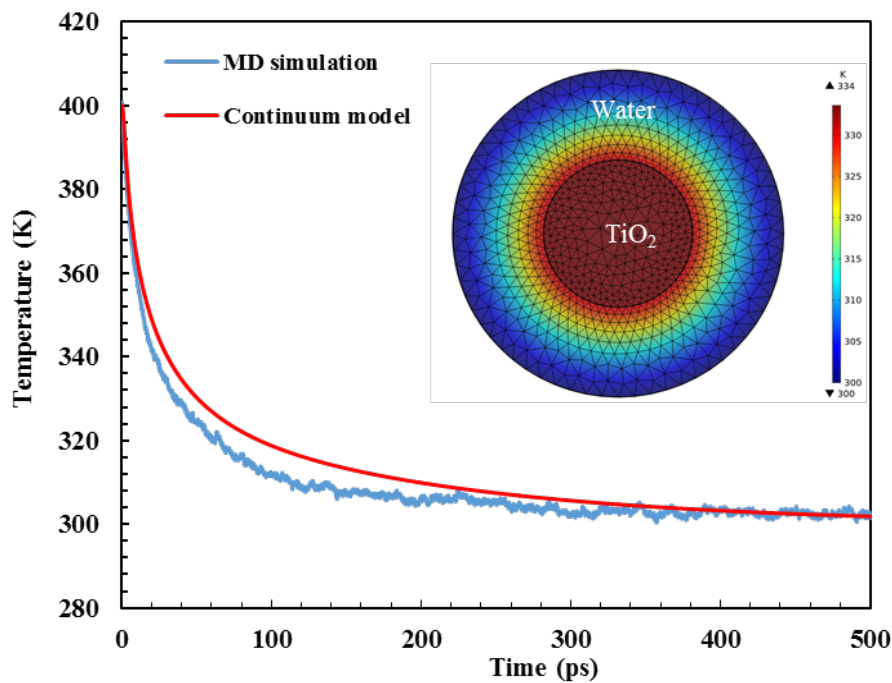

Fig. 11. Temperature of TiO$_2$ nanoparticle during the thermal relaxation time according to MD simulation and continuum conduction model. The temperature of water at distance of 10 nm from the center of TiO$_2$ nanoparticle is assumed to be 300 K.

## 7. Conclusion

In this study, MD simulation was utilized to investigate the effects of temperature and diameter of TiO$_2$ nanoparticle, and surface wettability on thermal conductance. Also, the influence of Coulombic interatomic potential on thermal conductance was studied. Moreover, a continuum model was introduced to investigate the thermal relaxation of a heated nanoparticle.

The acquired findings are summarized as follows:

a) The interfacial thermal conductance between TiO$_2$ and water is about one order of magnitude higher than that of water immersed gold, silver, platinum, copper, silicon nanoparticles or graphene flakes and CNTs.
b) By increasing the diameter of TiO$_2$ from 4 to 9 nm, the interfacial thermal conductance between nanoparticle and base fluid decreases slightly.
c) Thermal conductance between TiO$_2$ nanoparticle and water raises by increasing the nanoparticle temperature in the heating-up stage.
d) When the interaction strength between TiO$_2$ nanoparticle and water increases, the interfacial thermal conductance enhances.
e) Coulombic interatomic potential has a significant effect on thermal conductance, so that, by not considering this potential energy from interaction between water and nanoparticle, the value of thermal conductance reduces about two orders of magnitudes.
f) The thermal relaxation time of a nanoparticle in liquid computed from MD simulation results are close to that calculated from a continuum conduction model.

The results of this study could play a role in the application of TiO$_2$ nanoparticles in nanofluids as well as their application in the treatment of cancer. Also, the effect of electrostatic interactions in significantly increasing the interfacial thermal conductance can lead to further elucidation of the role of electrical charges loaded on the neutral particles to enhance the heat transfer from nanoparticles to the surrounding water.

**Data availability**

The data that support the findings of this study are available from the corresponding author, upon reasonable request.